\documentclass[reqno, letterpaper, 12pt]{amsart}
\usepackage{amsmath}
\usepackage{amsthm}
\usepackage{amssymb}
\usepackage{subfig}
\usepackage{graphicx}
\usepackage[unicode]{hyperref}
\usepackage{accents}
\usepackage{color}
\usepackage{placeins}
\usepackage{setspace}
\usepackage{lineno}
\usepackage{amsbsy}
\usepackage[usenames,dvipsnames]{pstricks}
\usepackage{epsfig}
\usepackage{pst-grad} % For gradients
\usepackage{pst-plot} % For axes
\usepackage{fullpage}
\numberwithin{equation}{section}

\title{A model for tracking fronts of stress-induced permeability enhancement}
\author{K.C. Lewis, Satish Karra and Sharad Kelkar}
\date{}
\address{Computational Earth Sciences Group\\
Earth and Environmental Sciences Division\\
 Los Alamos National Laboratory (LANL)\\
Los Alamos, NM 87545\\
United States of America}
\email{kaylal@lanl.gov (K.C.Lewis);}
\email{satkarra@lanl.gov (S. Karra);}
\email{kelkar@lanl.gov (S. Kelkar)}

\begin{document}

\thanks{Corresponding Author: K.C. Lewis, 
Computational Earth Sciences Group,
Earth and Environmental Sciences Division,
 Los Alamos National Laboratory (LANL),
Los Alamos, NM 87545,
United States of America. Phone: 505-665-6803 Email: kaylal@lanl.gov}

\tableofcontents
\newpage

\doublespacing

\begin{abstract}
Using an analogy to the classical Stefan problem, we construct evolution equations for the fluid pore pressure on both sides of a propagating stress-induced damage front. Closed form expressions are derived for the position of the damage front as a function of time for the cases of thermally-induced damage as well as damage induced by over-pressure. We derive expressions for the flow rate during constant pressure fluid injection from the surface corresponding to a spherically shaped subsurface damage front. Finally, our model results suggest an interpretation of field data obtained during constant pressure fluid injection over the course of 16 days at an injection site near Desert Peak, NV.
\end{abstract}

\maketitle

\linenumbers
\section{Introduction} \label{Intro}
The formation and propagation of subsurface stress-induced damage zones is of great practical interest for oil extraction, geothermal energy, and CO$_2$ sequestration (e.g., \cite{Rutqvist}, \cite{Yow}, \cite{Kohl}, \cite{Dusseault}). For all three applications, it is desirable to understand how human activities may affect permeability in the subsurface. In order to achieve this goal, it is important to understand a complex array of interrelated factors including local geological features, the in situ stress state, and which of several modes of stress-induced failure are most likely to dominate in a given scenario.\par
Two commonly employed mechanical failure criteria are the tensile and Mohr-Coulomb criteria. The type of rock failure likely to occur is governed by the conditions of in situ stress and the pressure of the fluid being injected. If the injection pressure is high enough to exceed the minimum principle in situ stress, a tensile ``hydraulic fracture'' is created. Walls of the fracture are pushed open by the fluid pressure creating a high permeability pathway through which the injected fluid can flow along the hydraulic fracture. The permeability of such an open fracture is commonly represented by a cubic law, where the permeability at any point in the fracture varies as the cube of the local fracture aperture and an empirically determined friction factor. Published results (\cite{LeeCho}) show that the aperture, and hence the permeability, of such a hydraulic fracture varies smoothly over the majority of the fracture length, dropping sharply to a very small value near the crack tip. For examples of analytic studies of tensile fracture propagation, see \cite{Geertsma}, \cite{Gordeyev}, and \cite{Wijesinghe}. \par

Rock failure can also occur at fluid pressures below the minimum principle earth stress through the mechanism of shearing. Such failure is often described using the Mohr-Coulomb criterion (for more detail, see \cite{JaegerCook}). The determining quantity in this case is the relative magnitude of the shear stress and the effective normal stress in the rock. The effective normal stress decreases as the fluid pressure increases. When the rock fails in shear, the fractures can dilate but do not display large aperture widening as in the case of hydraulic fractures, although a significant increase in permeability does take place in the plane of failure (\cite{Lockner}, \cite{MitchellFaulkner}). The situation is more complex than that of a fracture open in tension, and there are no simple analytical expressions relating fracture geometry to permeability that are widely applicable. Published results (\cite{LeeCho}) show an increase in permeability by factors up to 100 or so under shear failure.\par

\subsection*{Main contributions}
A complete analysis of the problem requires solving the coupled nonlinear equations of fluid flow, heat transfer, and mechanical deformation of the rock mass, necessitating the use of numerical models. However, useful insight into the behavior of the system can be obtained using simplified conceptual models that allow the system of governing equations to be decoupled. We solve the fluid flow problem, while incorporating the influence of the mechanical deformation and rock failure aspects implicitly through a prescribed step change in the permeability, with a low permeability for rock in an undamaged state and a higher permeability for rock that is in a fully damaged state. That is, we conceptualize the system as comprised of two zones - a zone containing the rock at pre-failure conditions and a second zone with post-failure conditions in the rock. Simplification is introduced by treating rock permeability and porosity as constant within each zone, with an abrupt change in the transition zone, which we approximate as infinitely thin. Further, we take the ratio of the post-failure to pre-failure permeability to be an empirically determined constant. We study two modes of failure. In the first, rock failure is driven by high fluid pressure, with a fixed specified pressure above which rock failure occurs and below which no damage occurs. We take this threshold pressure to be constant, which assumes an isothermal environment as well as an approximately uniform initial stress field. In the second case, rock failure is driven thermally due to a large temperature difference between fluid and the surrounding medium (see, e.g., \cite{AharonovAnders} and \cite{VoightElsworth}). While in this case there is no assumed threshold pressure, our analysis does assume that differences in stress between one side of a propagating damage front and the other are approximately constant in time as the front propagates.\par

Using these assumptions we derive approximate expressions for the position of the interface between damaged and undamaged regions during failure of a porous matrix induced by fluid injection. Afterward, we show how to relate these expressions to the mass flow rate during constant pressure injection, which is a commonly employed observable quantity. Finally, we show how our model leads to one plausible interpretation of flow rate data obtained during constant pressure injection at a site near Desert Peak, Nevada. The following are the main contributions of this paper:
\begin{enumerate}
\item The model we present can be used to predict the position of the damage front without explicitly solving the coupled equations governing stress and fluid flow.
\item The model relates subsurface damage to mass flow rates during constant pressure fluid injection.
\item The model includes no assumption regarding whether failure occurs as a result of tension or shear.
\item The model is a new application of the Stefan-type solution method.
\end{enumerate}

\subsection*{Outline}
In section \ref{StefanAnalogy} we first draw an analogy with the Stefan problem and present the governing equations. Then we present an analytical expression for the position of, and pore pressures on either side of, a vertical damage front; we then derive an approximate closed form expression for the position of the front in section \ref{FrontPosition1}. Next, in section \ref{HorizFront}, we show that the effect of gravity is small for sufficiently small times. We then adapt our model to calculate the approximate damage front position in the case of thermally driven failure. Expressions for flow rates under constant pressure injection for spherical damage front geometry are derived in section \ref{FlowRates} followed by comparison of the closed form solutions from our model with the field data from Desert Peak, Nevada in section \ref{Disc}.

\section{Analytical model} \label{AnalyticModel}
\subsection{Analogy with the Stefan problem} \label{StefanAnalogy}
Consider a semi-infinite horizontal slab of fully saturated porous material with a high pore pressure, $p_H$, maintained at the left end and such that, at all times, the pressure approaches a much lower pressure $p_L$ as $x$ approaches infinity. The initial pressure is $p_L$ everywhere, and material properties are initially uniform. Following the discussion of last section, we assume that there is a sharp boundary separating damaged and undamaged material. At times $t > 0$ a damage front will travel in the positive $x$ direction, and all positions to the left of the front will be in the ``damage'' zone. On the other hand, all points to right of this point will have their initial permeabilities and porosities. The permeability is clearly a function of pressure, being a higher value on the left side of the damage front than on the right side; this pressure dependence causes the mass balance equation (see below) to become nonlinear. However, the constancy of material properties on either side of the damage front motivates the idea of splitting the solution domain into two halves, solving linear mass balance equations on each half, and pasting the resulting solutions together at the damage front. This situation is exactly analogous to that in the classical Stefan problem (see \cite{HeatConduct}, \cite{Rubinstein}).\par

In one version of the classical Stefan problem, the half plane corresponding to $x \leq 0$ is filled with ice while that corresponding to $x > 0$ is filled with liquid water. As time progresses, an ice front propagates toward positive values of $x$ as the ice phase removes heat from the liquid. The problem is to solve for both the temperature as a function of time and space for all $x > 0$ and the position of the ice-water interface as a function of time. Mathematical problems of a similar type arise in the study of systems that have moving boundaries separating regions with distinct physical properties (for many such examples, see \cite{Rubinstein}).\par 
\subsection{Governing equation} \label{GovEqus}
\begin{figure} \centering
\includegraphics[width=0.7\textwidth]{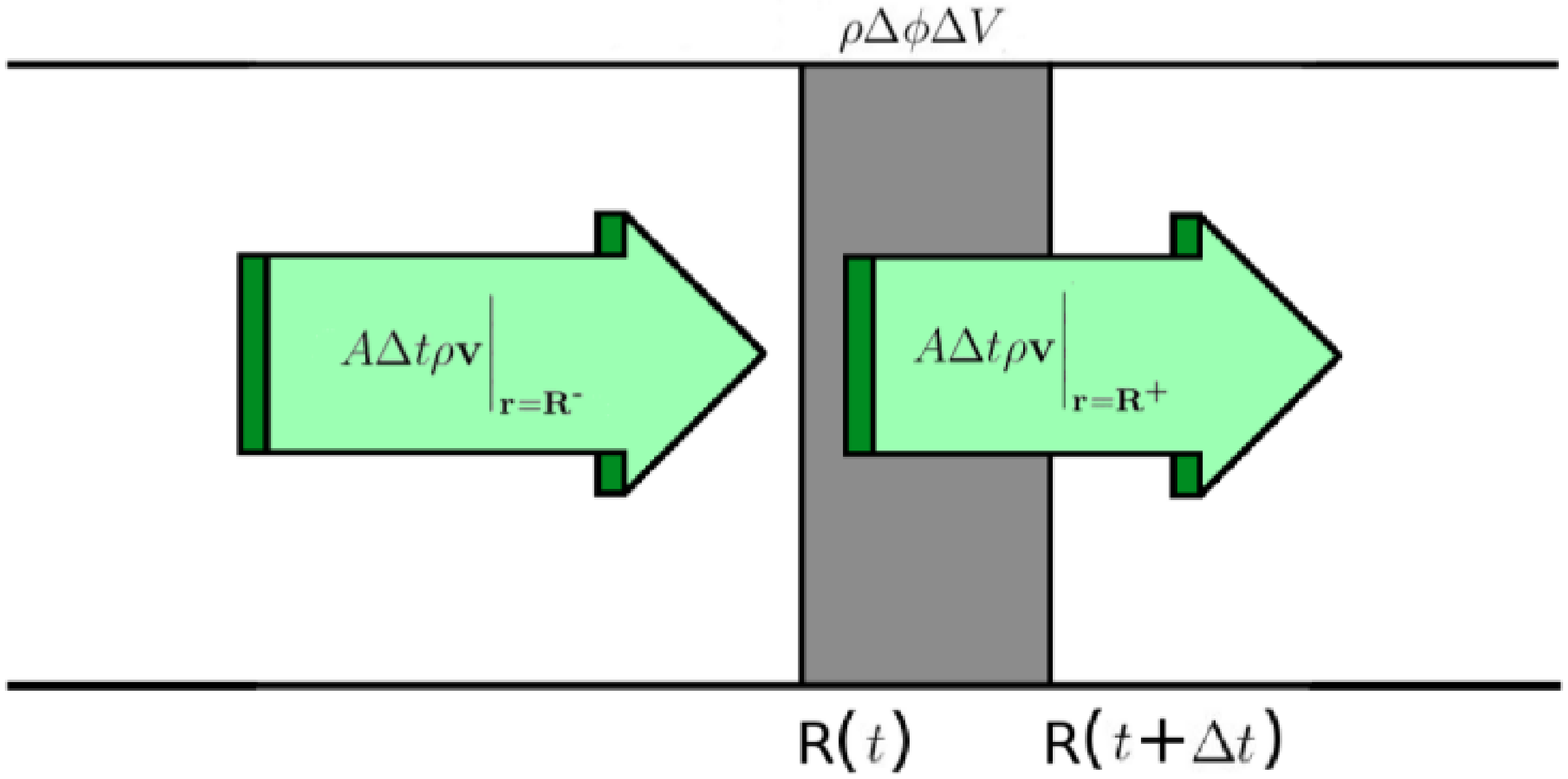}
\caption{Representation of the mass balance condition (\ref{StefanCond1}). The mass of fluid flowing toward the damage front from the side of the damaged material equals the mass of fluid flowing away from the front into the undamaged material, plus the fluid taken up due to an increase in porosity as the front traverses a volume $\Delta V$ in time $\Delta t$.} \label{StefanCondition}
\end{figure}

On each side of the damage front the permeability and porosity are taken as constants, but such that each displays a discrete jump in crossing from one side of the front to the other. The fluid on each side of the damage front therefore obeys a mass conservation equation
\begin{linenomath*} \begin{equation} \label{MassConserve1}
\frac{\partial (\rho \phi_i)}{\partial t}+\nabla \cdot (\rho \textbf{v}_i) = 0,
\end{equation} \end{linenomath*}
where $\rho$ is the fluid density, $\phi$ is the porosity, $\textbf{v}$ is the volumetric flux, and the subscript $i=1,2$, represents the damaged or undamaged side of the front. The volumetric flux is given by Darcy's law
\begin{linenomath*} \begin{equation} \label{VolFlux}
\textbf{v}_i = -\frac{k_i}{\mu}\left( \nabla p_i + \rho g \nabla z \right),
\end{equation} \end{linenomath*}
where $k$ is the permeability, $\mu$ is the dynamic viscosity, $p$ is the pressure, $g$ is the gravitational acceleration, and $z$ is the vertical coordinate taken as positive upward. We include variations in the fluid density only in the unsteady term and neglect the gradient of the fluid density (see the Appendix for a detailed justification of this assumption). On each side of the front, the porosity and the density are related to the pressure via
\begin{linenomath*} \begin{equation} \label{PorosPres}
\phi_i = \phi_{i0} + \alpha(p_i - p_0),
\end{equation} \end{linenomath*}
and
\begin{linenomath*} \begin{equation} \label{DensPres}
\rho = \rho_0 [1 + \beta (p_i - p_0)],
\end{equation} \end{linenomath*}
where $\alpha$ is a constant, $\beta$ is the fluid compressibility, and the subscript zero refers to initial values. The reasoning leading to (\ref{PorosPres}) can be found in \cite{LewisArctic}. The increase in $\phi$ in crossing from the undamaged to the damaged side of the front is assumed constant and equal to $\Delta \phi \equiv \phi_1 - \phi_2$. Combining (\ref{MassConserve1}) through (\ref{DensPres}), we obtain
\begin{linenomath*} \begin{equation} \label{MassConserve2}
\frac{\partial p_i}{\partial t} - a_i \nabla^2 p_i = 0,
\end{equation} \end{linenomath*}
where
\begin{linenomath*} \begin{equation} \label{DefinitionDiffuse}
a_i \equiv \frac{k_i}{\mu(\phi_i\beta+\alpha)} \equiv \frac{k_i}{\mu \gamma_i},
\end{equation} \end{linenomath*}
and where $\gamma_i$ is the total compressibility (liquid plus porous medium) in region $i$. We impose the boundary conditions
\begin{linenomath*} \begin{equation} \label{BoundaryConds}
\begin{array} {lr} p_1(\textbf{\textbf{r} = 0},t) = p_H, & \\ \\ p_2(\textbf{r} \rightarrow \infty,t) = p_L, \\ \\ p_1(\textbf{\textbf{r} = R},t)=p_2(\textbf{\textbf{r} = R},t)=p_D, & \end{array}
\end{equation} \end{linenomath*}
where $\textbf{R}(t)$ is the position of the damage zone at time $t$, $p_2(\textbf{r} \rightarrow \infty,t)$ is an abbreviation for the value of $p_2$ as $|\textbf{r}|$ approaches infinity, and $p_D$ is defined as the pore pressure at the damage front. For uniqueness of the solution, one more boundary condition must be imposed at the damage front. Over an increment of time, the fluid mass into the interior (damaged) side of the damage front must equal that out of the exterior (undamaged) side minus the amount of fluid taken up by an increase in pore volume due to progression of the front. Requiring mass conservation across an element of area $A$ of damage front over a time $\Delta t$ thus yields the boundary condition
\begin{linenomath*} \begin{equation} \label{StefanCond1}
A \Delta t \rho \textbf{v}\bigg|_{\textbf{r} = \textbf{R}^\textbf{-}} \cdot \textbf{\^{n}} = A \Delta t \rho \textbf{v} \bigg|_{\textbf{r}=\textbf{R}^\textbf{+}} \cdot \textbf{\^{n}} + \rho \Delta \phi \Delta V,
\end{equation} \end{linenomath*}
where $\Delta V$ is the total volume traversed by the front over $\Delta t$, $\Delta \phi \equiv \phi_1 - \phi_2 > 0$, and \textbf{\^{n}} is the unit normal to $A$ (see figure \ref{StefanCondition}). In (\ref{StefanCond1}) we have neglected variations in fluid density due to progression of the damage front; this we justify in section (\ref{NeglectDens}). Substituting (\ref{VolFlux}) into (\ref{StefanCond1}) gives the final boundary condition as
\begin{linenomath*} \begin{equation} \label{StefanCond2}
k_1\left( \nabla p_1 + \rho g \nabla z \right) \cdot \textbf{\^{n}} = k_2\left( \nabla p_2 + \rho g \nabla z \right) \cdot \textbf{\^{n}} - \mu \Delta \phi \frac{1}{A}\frac{dV}{dt}.
\end{equation} \end{linenomath*}
\subsection{Solution for a vertical planar damage front} \label{VertFront}
If the damage front is assumed to be a vertical plane, and if the fluid flux parallel to the plane of the front is negligible compared to the flux normal to the plane of damage, then (\ref{MassConserve2}) becomes
\begin{linenomath*} \begin{equation} \label{DiffusionEqu}
\frac{\partial p_i}{\partial t} - a_i \frac{\partial^2 p_i}{\partial x^2} = 0,
\end{equation} \end{linenomath*}
and equation (\ref{StefanCond2}) becomes
\begin{linenomath*} \begin{equation} \label{StefanCondHoriz}
k_1 \frac{\partial p_1}{\partial x}\bigg|_{x=X} = k_2 \frac{\partial p_2}{\partial x}\bigg|_{x=X} - \mu \Delta \phi \frac{dX}{dt},
\end{equation} \end{linenomath*}
where the damage front is located at $x=X$. The partial differential equation plus boundary and initial conditions given above can be solved exactly as in \cite{HeatConduct}, pg. 285, with the substitutions $p \mapsto v, p_H \mapsto 0, p_D \mapsto T_1, p_L \mapsto V, a \mapsto \kappa, k \mapsto K$, and $\mu \Delta \phi \mapsto L \rho$; however, for the conveniance of the reader we now briefly recapitulate the argument leading to a solution.\par
Scale analysis suggests that the solution to (\ref{DiffusionEqu}) depends only on the dimensionless combination $x/\sqrt{a_i t}$ (see \cite{Barenblatt}). Substituting $p$ as a function of $x/\sqrt{a_i t}$ into (\ref{DiffusionEqu}) results in an ordinary differential equation that can be easily integrated to give the solution
\begin{linenomath*} \begin{equation} \label{1DStefanSoln}
p_i(x,t) = \frac{2 C_i}{\sqrt{\pi}} \int_0^{x/\sqrt{t}} e^{-\zeta^2/4a_i} d\zeta + D_i = C_i \mathrm{erf}\left(\frac{x}{2\sqrt{a_i t}}\right) + D_i,
\end{equation} \end{linenomath*}
where $\mathrm{erf}$ is the error function, defined as
\begin{linenomath*} \begin{equation} \label{ErrFunc}
\mathrm{erf}(x) \equiv \frac{2}{\sqrt{\pi}}\int_0^x e^{-z^2}dz. 
\end{equation} \end{linenomath*}
The first two boundary conditions from (\ref{BoundaryConds}) yield
\begin{linenomath*} \begin{equation} \label{ConstantsD}
\begin{array} {lr} D_1 = p_H, \\ \\ D_2 = p_L - C_2, & \end{array}
\end{equation} \end{linenomath*}
so that it only remains to find the constants $C_1$ and $C_2$. The third of equations (\ref{BoundaryConds}) yields
\begin{linenomath*} \begin{equation} \label{Continuity}
C_1 \mathrm{erf}\left(\frac{X}{2\sqrt{a_1 t}}\right)+p_H = C_2\left[ \mathrm{erf}\left(\frac{X}{2\sqrt{a_2 t}}\right)-1\right]+p_L=p_D.
\end{equation} \end{linenomath*}
The first and middle expressions can only be equal to the constant on the right if $X=\lambda\sqrt{t}$ for some constant $\lambda$. Substituting this expression for $X$ into (\ref{Continuity}) allows one to solve for both $C_1$ and $C_2$ as functions of the undetermined constant $\lambda$. Application of condition (\ref{StefanCondHoriz}) then results in the equation
\begin{linenomath*} \begin{equation} \label{ImplicitLambda}
\frac{k_1 C_1(\lambda)e^{\frac{-\lambda^2}{4a_1}}}{\sqrt{a_1}} = \frac{k_2 C_2(\lambda)e^{\frac{-\lambda^2}{4a_2}}}{\sqrt{a_2}} - \frac{\mu \Delta \phi \lambda \sqrt{\pi}}{2},
\end{equation} \end{linenomath*}
which determines $\lambda$ implicitly. In general, equation (\ref{ImplicitLambda}) can be solved for $\lambda$ only numerically in combination with the constraints on $C_1$ and $C_2$ from equations (\ref{Continuity}); however, in the next section we show how to obtain an approximate closed form expression for $X = \lambda \sqrt{t}$. 
\subsection{Approximate Expression for the damage front position} \label{FrontPosition1}
\begin{figure} \centering
\includegraphics[width=0.7\textwidth]{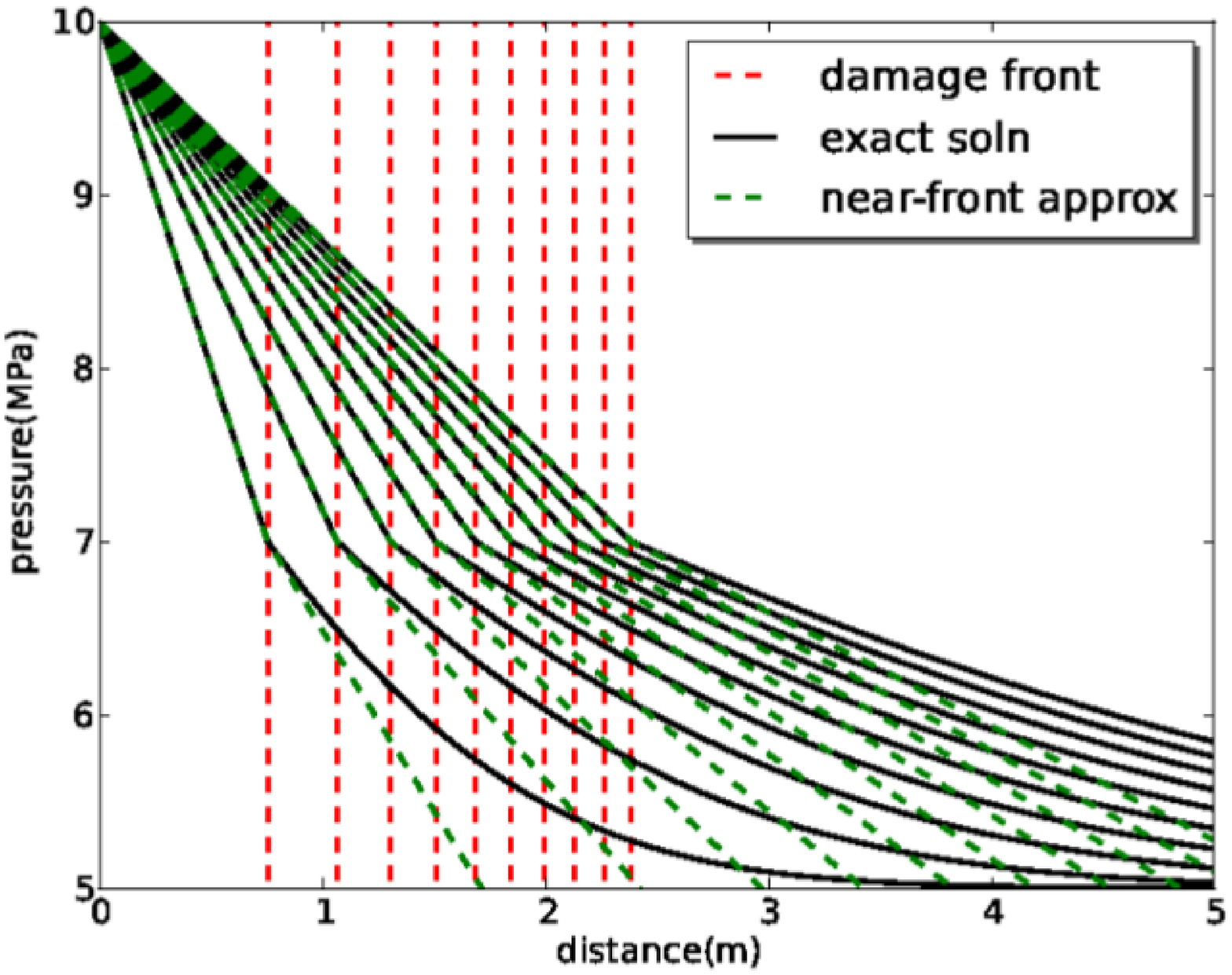}
\caption{Representative pressure profiles calculated from (\ref{1DStefanSoln}) using $\mu = 10^{-4}$ Pa$\cdot$s, $\Delta \phi = 0.01$, $\gamma = 10^{-10}$ Pa$^{-1}$, $k_1 = 10^{-13}$ m$^2$, $k_2 = 10^{-14}$ m$^2$, $p_H = 10$ MPa, $p_L = 5$ MPa, and $p_D = 7$ MPa. The profiles correspond to times of 1, 2, ..., 10 days. Red dashed lines indicate the distance of the damage front from the injection well. Green dashed lines show the approximate pressures used to estimate the derivative of the pressure directly adjacent to the damage front on each side. These lines lie directly on top of the exact solution in the zone of failed material.} \label{1D-Pressures}
\end{figure}

In the absence of any damage, i.e., $a_1=a_2\equiv a$, the effect of the high pressure at $x=0$ is governed by (\ref{DiffusionEqu}) and will travel a distance $L$ in time $t$ given approximately by the characteristic diffusive length scale
\begin{linenomath*} \begin{equation}
L = \sqrt{a t}.
\end{equation} \end{linenomath*}
In fact, the form of this length scale does not depend on the problem geometry - it depends only on the fact that the relevant process is one of diffusion (\cite{Barenblatt}, \cite{HeatConduct}). In the case of a propagating damage front, the pressure at the $x=0$ boundary has influenced that at the damage front, by definition, enough to raise the pressure there to $p_D$. Furthermore, the speed at which an effect from the high pressure boundary can propagate is limited by the lower permeability of the undamaged region as well as by increased fluid storage due to the porosity increase upon damage. Therefore, the diffusive time scale for region one is short compared to that governing the movement of the damage front; this fact implies that the pressure in region one at all times assumes approximately a linear steady-state profile with gradient
\begin{linenomath*} \begin{equation} \label{ApproxGradient1}
\frac{\partial p_1}{\partial x} \approx \frac{p_D - p_H}{X} \equiv -\frac{\Delta p_1}{X}.
\end{equation} \end{linenomath*}
This approximation improves as $p_D$ approaches $p_H$. In region two, the pressure effect from the damage front propagates to roughly the distance $\sqrt{a_2 t}$ in time $t$. Therefore, an approximation similar to (\ref{ApproxGradient1}), using the distance $\sqrt{a_2 t}$ instead of $X$, can be used to estimate $\partial p_2/\partial x$; the approximate pressure gradient in region two is given as
\begin{linenomath*} \begin{equation} \label{ApproxGradient2}
\frac{\partial p_2}{\partial x} \approx \frac{p_L - p_D}{\sqrt{a_2 t}} \equiv -\frac{\Delta p_2}{\sqrt{a_2 t}}.
\end{equation} \end{linenomath*}
See Figure \ref{1D-Pressures} for a comparison between (\ref{ApproxGradient1}) and (\ref{ApproxGradient2}) and the exact solution slopes given by (\ref{1DStefanSoln}). Putting (\ref{ApproxGradient1}) and (\ref{ApproxGradient2}) into (\ref{StefanCondHoriz}) leads to 
\begin{linenomath*} \begin{equation} \label{ApproxStefanCond}
\frac{k_1 \Delta p_1}{X} = \frac{k_2 \Delta p_2}{\sqrt{a_2 t}} + \mu \Delta \phi \frac{dX}{dt}.
\end{equation} \end{linenomath*}
We search for a solution of the form $X = \lambda t^n$ for some undetermined $n$. Putting this expression into (\ref{ApproxStefanCond}) yields
\begin{linenomath} \begin{equation}
\frac{k_1\Delta p_1}{\lambda}t^{-n} - \frac{k_2\Delta p_2}{\sqrt{a_2}}t^{-1/2} - \mu\Delta \phi \lambda n t^{n-1} = 0.
\end{equation} \end{linenomath}
The only way that this equation can be satisfied for all times is for the powers of $t$ to equal one another; the only value of $n$ for which such is the case is $n=1/2$. $X$ therefore takes the form $\lambda \sqrt{t}$ and (\ref{ApproxStefanCond}) becomes
\begin{linenomath*} \begin{equation}
\lambda^2 + \left( \frac{2k_2\Delta p_2}{\mu \Delta \phi \sqrt{a_2}} \right) \lambda - \frac{2k_1\Delta p_1}{\mu \Delta \phi} = 0.
\end{equation} \end{linenomath*}
There is only one positive root of this equation, leading to the approximate damage front position
\begin{linenomath*} \begin{equation} \label{ApproxFrontPosition}
X = \left( -\frac{k_2\Delta p_2}{\mu \Delta \phi \sqrt{a_2}} + \sqrt{\frac{k_2^2\Delta p_2^2}{\mu^2 \Delta \phi^2 a_2} + \frac{2k_1\Delta p_1}{\mu \Delta \phi}} \right) \sqrt{t}.
\end{equation} \end{linenomath*}
Table \ref{LambdaCompare} shows values of $\lambda$ calculated from (\ref{ApproxFrontPosition}) and values computed numerically from equation (\ref{ImplicitLambda}) via the bisection method for a wide range of permeabilities and porosities for the damaged and undamaged zones. Every row in the table corresponds to $p_H = 3$ MPa, $p_L = 0.1$ MPa, $p_D = 1.5$ MPa, $\mu = 10^{-3}$ Pa $\cdot$ s, and $\gamma = 10^{-10}$ Pa$^{-1}$, but the results are not very sensitive to changes in these parameters. We note that the relative error with respect to the computationally derived value of $\lambda$ does not exceed three percent. The largest relative errors occur when flow on the exterior side of the damage front is largest, because approximation (\ref{ApproxGradient2}) is not as good an approximation as (\ref{ApproxGradient1}).\par
The expression for $\lambda$ can be further simplified if there is a large contrast in the porosity and permeability on crossing from one side of the damage front to the other. To affect the simplification, we first re-write the approximate expression for $\lambda$ as
\begin{linenomath*} \begin{equation} \label{ApproxLambda}
\lambda \approx -\frac{k_2 \Delta p_2}{\mu \Delta \phi \sqrt{a_2}}+\sqrt{\frac{2k_1\Delta p_1}{\mu \Delta \phi}}\sqrt{\frac{k_2^2 \Delta p_2^2}{2 k_1 \Delta p_1 \mu \Delta \phi a_2} + 1}.
\end{equation} \end{linenomath*}
The second term on the right side is greater than
\begin{linenomath*} \begin{equation} \label{RegionOneTerm}
\sqrt{\frac{2k_1\Delta p_1}{\mu \Delta \phi}},
\end{equation} \end{linenomath*}
so if the absolute value of the first term on the right side of (\ref{ApproxLambda}) is much less than this quantity, it may be neglected. This condition may be written as
\begin{linenomath*} \begin{equation}
\frac{k_2^2 \Delta p_2^2}{\mu^2 \Delta \phi^2 a_2} << \frac{2 k_1 \Delta p_1}{\mu \Delta \phi},
\end{equation} \end{linenomath*}
which is completely equivalent to
\begin{linenomath*} \begin{equation} \label{ApproxCondition1}
\frac{k_2^2 \Delta p_2^2}{2 k_1 \Delta p_1 \mu \Delta \phi a_2} << 1.
\end{equation} \end{linenomath*}
Therefore, if (\ref{ApproxCondition1}) holds, the first term on the right side of (\ref{ApproxLambda}) may be neglected. But (\ref{ApproxCondition1}) is also the condition that the factor multiplying (\ref{RegionOneTerm}) in equation (\ref{ApproxLambda}) is approximately equal to unity. Satisfaction of condition (\ref{ApproxCondition1}) therefore results in
\begin{linenomath*} \begin{equation} \label{SimpleLambda}
\lambda \approx \sqrt{\frac{2k_1\Delta p_1}{\mu \Delta \phi}}.
\end{equation} \end{linenomath*}
Condition (\ref{ApproxCondition1}) can be made more transparent by using (\ref{DefinitionDiffuse}) to eliminate $a_2$ and assuming that $\Delta p_2 \approx \Delta p_1$. Then (\ref{ApproxCondition1}) takes the form 
\begin{linenomath*} \begin{equation} \label{SmallPermCond}
\frac{\gamma_2k_2\Delta p_1}{2 k_1 \Delta \phi} << 1.
\end{equation} \end{linenomath*}
Hence, if the contrast in material properties between regions one and two is large enough to satisfy (\ref{SmallPermCond}), equation (\ref{SimpleLambda}) may be employed to estimate the position of the damage front as $X \approx \lambda \sqrt{t}$. Equation (\ref{SimpleLambda}) is the same expression that would have been obtained if flow across the damage front had been neglected in equation (\ref{ApproxStefanCond}), i.e., if the term involving $\partial p_2/\partial x$ had been neglected. Therefore, condition (\ref{SmallPermCond}) is also the condition that flow across the damage front toward the lower permeability region may be neglected in determining the position of the front. As an example, if $\gamma_2 = 10^{-10}$ Pa$^{-1}$, $k_2 = 10^{-16}$ m$^2$, $\Delta p_1 = 10^6$ Pa, $k_1 = 10^{-14}$ m$^2$, and $\Delta \phi = 0.1$, the quantity on the left hand side of (\ref{SmallPermCond}) is equal to $0.5 \times 10^{-5}$. 
\subsection{Effect of gravity on the damage front position} \label{HorizFront}
In the previous section we assumed that the damage front is a vertical planar surface; gravity did not appear in the boundary or initial conditions because fluid flow in the vertical direction was assumed negligible compared to that in the horizontal direction. We now consider the case such that the damage front is a horizontal planar surface and vertical fluid flow dominates. The presence of gravity in the volumetric flux gives rise to a boundary condition that prevents the method of solution employed section \ref{VertFront}; however, it is still possible to derive an approximate formula for the position of the damage front. \par
In the present case, equation (\ref{StefanCond2}) becomes 
\begin{linenomath*} \begin{equation} \label{StefanCondVert}
k_1\left( \frac{\partial p_1}{\partial z} + \rho g \right) = k_2\left( \frac{\partial p_2}{\partial z} + \rho g \right) - \mu \Delta \phi \frac{dZ}{dt}.
\end{equation} \end{linenomath*}
If condition (\ref{SmallPermCond}) holds, we may neglect flow across the damage front. Then, by using (\ref{ApproxGradient1}), equation (\ref{StefanCondVert}) may be written in the form
\begin{linenomath*} \begin{equation} \label{NonSeparableEqu}
\frac{\Delta p_1}{\Delta p_2} \approx \Delta \phi \left(\frac{\mu Z}{k_1 \Delta p_2}\frac{dZ}{dt}\right) + \frac{\rho g Z}{\Delta p_2}.
\end{equation} \end{linenomath*}
This equation cannot be easily integrated, but a useful solution can still be obtained by noting that the second term on the right hand side is small relative to unity when
\begin{linenomath*} \begin{equation}
\frac{\rho g}{\Delta p_2} << \frac{1}{Z}.
\end{equation} \end{linenomath*}
For typical orders of magnitude of the quantities on the left hand side, this inequality becomes $Z << 100$ m. When this condition holds, we may take $\epsilon \equiv \rho g / \Delta p_2$ as a small parameter. The solution may then be represented as a perturbative series
\begin{linenomath*} \begin{equation} \label{PerturbSeries}
Z(t) = \sum_{n=0}^\infty Z_n(t) \epsilon^n. 
\end{equation} \end{linenomath*}
Substituting (\ref{PerturbSeries}) into (\ref{NonSeparableEqu}), setting coefficients of differing powers of $\epsilon$ equal to zero, and neglecting powers of $\epsilon$ greater than unity yields the equations
\begin{linenomath*} \begin{equation} \label{ZerothOrderEqu}
\frac{\Delta \phi \mu}{k_1 \Delta p_1} Z_0 \frac{dZ_0}{dt} = 1,
\end{equation} \end{linenomath*}
and
\begin{linenomath*} \begin{equation} \label{FirstOrderEqu}
\frac{dZ_1}{dt} + \frac{Z_1}{2t} = -\frac{k_1 \Delta p_1}{\Delta \phi \mu}.
\end{equation} \end{linenomath*}
This equation is dimensionally homogeneous because $Z_1$ has dimensions of length squared, due to $\epsilon$ having dimensions of $1/$length. The initial condition for these equations is $Z_{0,1}(0) = 0$. Equation (\ref{ZerothOrderEqu}) has the solution
\begin{linenomath*} \begin{equation}
Z_0 = \sqrt{\frac{2 k_1 \Delta p_1}{\mu \Delta \phi} t}.
\end{equation} \end{linenomath*}
Equation (\ref{FirstOrderEqu}) can be easily integrated to give
\begin{linenomath*} \begin{equation}
Z_1 = -\frac{2k_1 \rho g t}{3 \Delta \phi \mu},
\end{equation} \end{linenomath*}
so that the perturbed solution to first order is
\begin{linenomath*} \begin{equation}
Z(t) \approx \sqrt{\frac{2 k_1 \Delta p_1}{\mu \Delta \phi} t} - \frac{2k_1 \rho g t}{3 \Delta \phi \mu}.
\end{equation} \end{linenomath*}
The ratio of the second term on the right hand side to the first is
\begin{linenomath*} \begin{equation}
\frac{\rho g}{3}\sqrt{\frac{2 k_1 t}{\mu \Delta p_1}},
\end{equation} \end{linenomath*}
and this term is small compared to unity for sufficiently small times. For example, if $k_1 = 10^{-13}$ m$^2$, the correction is small for times that are small compared to ten days. The effect of gravity is to slow the progression of an upward moving front, and this effect is more pronounced as $t$, or equivalently $Z$, increases (``equivalently'' because $Z$ is monotonically increasing in $t$).\par

\subsection{Damage for spherical geometry}
In spherical coordinates, the steady-state solution to (\ref{MassConserve2}) does not have the simple linear profile employed above; therefore, we separately derive a formula for the approximate damage front position in spherical geometry.
In the steady state and in spherical coordinates with radial symmetry, equation (\ref{MassConserve2}) becomes
\begin{linenomath*} \begin{equation} \label{SphericalMassConserve}
\frac{d^2 (rp_i)}{d r^2} = 0.
\end{equation} \end{linenomath*}
The solution to this equation is readily found to be
\begin{linenomath*} \begin{equation} \label{SphericalGenSoln}
p_i(r) = C_1 + \frac{C_2}{r},
\end{equation} \end{linenomath*}
where $C_1$ and $C_2$ are constants. The pressure profiles in the damaged and undamaged zones are then approximately (using reasoning similar to that in section \ref{FrontPosition1})
\begin{linenomath*} \begin{equation} \label{SphrPresLeft}
p_1(r) = p_H - \frac{R\Delta p_1}{R - r_0}\left(1-\frac{r_0}{r}\right),
\end{equation} \end{linenomath*}
and
\begin{linenomath*} \begin{equation} \label{SphrPresRight}
p_2(r) = p_D - \frac{\Delta p_2 \sqrt{a_2 t}}{\sqrt{a_2 t}-R}\left( 1 - \frac{R}{r} \right),
\end{equation} \end{linenomath*}
where $r_0$ is the radius of the injection well, i.e., $p_1(r_0) = p_H$.
Using these expressions, condition (\ref{StefanCond2}) becomes
\begin{linenomath*} \begin{equation} \label{StefanCondSphr}
\frac{k_1\Delta p_1 r_0}{R(R - r_0)} = \frac{k_2\Delta p_2 \sqrt{a_2 t}}{R(\sqrt{a_2 t}-R)} + \mu\Delta\phi\frac{dR}{dt}.
\end{equation} \end{linenomath*}
To affect a solution, we consider the case where $k_2/k_1 \ll 1$ and $r_0/R \ll 1$. Then (\ref{StefanCondSphr}) becomes
\begin{linenomath*} \begin{equation}
\frac{k_1\Delta p_1 r_0}{\mu \Delta \phi}=R^2\frac{dR}{dt},
\end{equation} \end{linenomath*}
which is separable and has the solution
\begin{linenomath*} \begin{equation} \label{SphrDamage}
R = \left( r_0^3 + \frac{3k_1\Delta p_1 r_0 t}{\mu \Delta \phi} \right)^\frac{1}{3}.
\end{equation} \end{linenomath*}
\subsection{Thermally induced damage} \label{ThermalDamage}
\begin{figure} \centering
\includegraphics[width=0.7\textwidth]{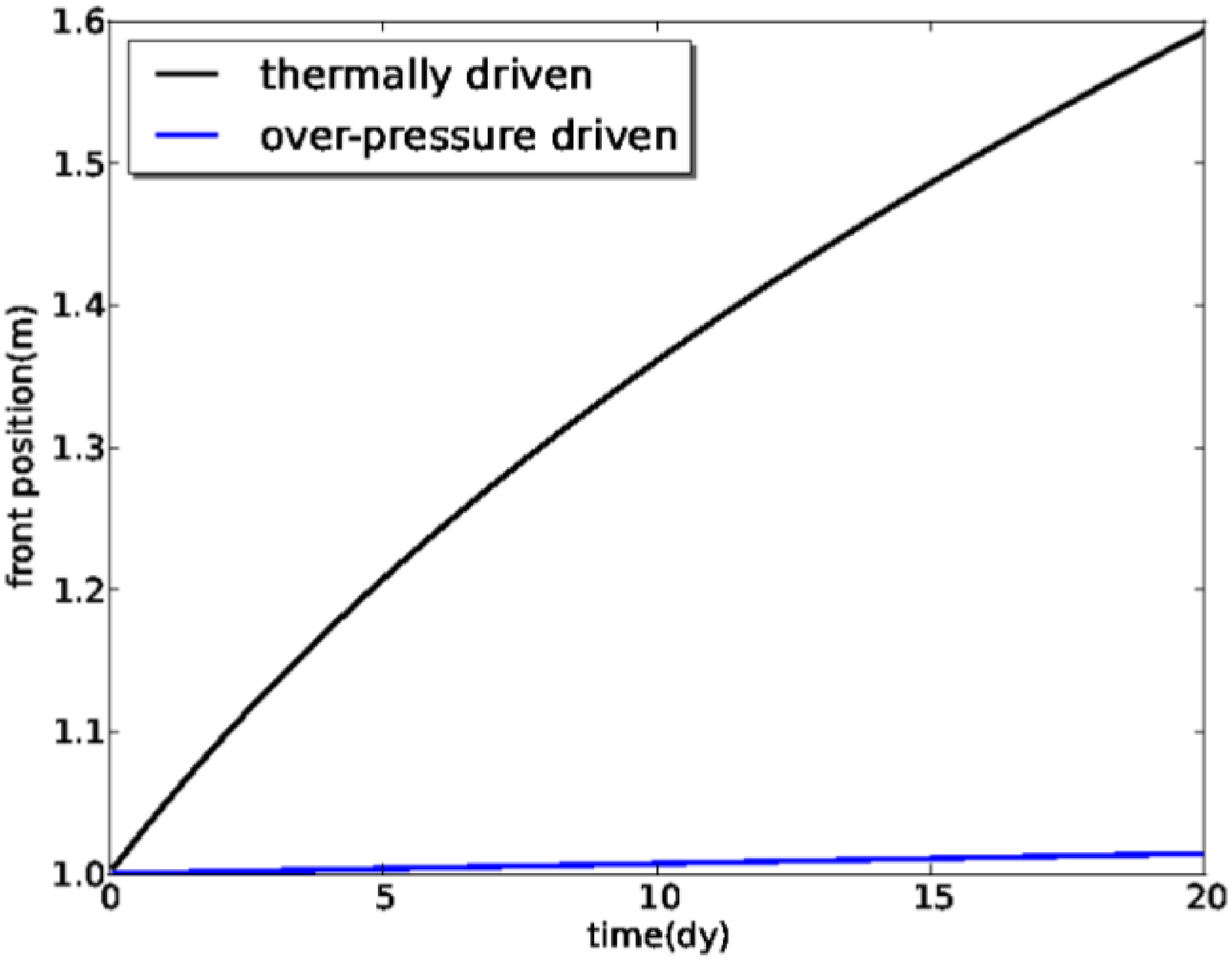}
\caption{Comparison between the positions of spherical damage fronts due to over-pressure (blue) and due to thermal effects (black).} \label{FrontCompare}
\end{figure}

When damage is driven by thermal effects rather than over-pressure, it is no longer reasonable to assume that the pressure at the damage front is approximately constant. We will now explore the consequences of letting $p_D$ vary, from $p_D = p_H$ when the front is at the injection source to $p_D = p_L$ as the front approaches infinity. The simplest assumption consistent with this behavior is that $p_H - p_D$ increases linearly with $R-r_0$. That is,
\begin{linenomath*} \begin{equation} \label{PressureDrop}
\Delta p_1 = \frac{\Delta p (R - r_0)}{R_{max} - r_0} \equiv D(R-r_0),
\end{equation} \end{linenomath*}
where $R_{max}$ is the distance at which $\Delta p_1 = p_H - p_\infty \equiv \Delta p$. In the following, we will only consider the system behavior for $r_0 < R_{max}$.
In the case of spherical geometry, we substitute (\ref{PressureDrop}) into (\ref{StefanCondSphr}) and again assume that $k_2/k_1 \ll 1$, obtaining
\begin{linenomath*} \begin{equation}
\frac{k_1Dr_0}{\mu \Delta \phi} = R\frac{dR}{dt},
\end{equation} \end{linenomath*}
which has the solution
\begin{linenomath*} \begin{equation} \label{ThermalSphrFront}
R = \sqrt{r_0^2 + \frac{2k_1\Delta pr_0 t}{\mu\Delta\phi(R_{max}-r_0)}}.
\end{equation} \end{linenomath*}
If $t_{max}$ is the time at which $R = R_{max}$, we may solve for $R_{max}$ in terms of this time as
\begin{linenomath*} \begin{equation} \label{MaxDistance}
R_{max} = \left(\frac{2 k_1\Delta p r_0 t_{max}}{\mu\Delta\phi}\right)^\frac{1}{3},
\end{equation} \end{linenomath*}
where we have assumed that $r_0/R_{max} \ll 1$. Figure \ref{FrontCompare} shows a comparison between front positions predicted via (\ref{ThermalSphrFront}) versus (\ref{SphrDamage}), using the parameters shown in Table \ref{FieldCompare} and $\Delta \phi = 10^{-2}$. \par
According to this model, then, the damage front progresses much faster in the case of thermally driven damage than in the case of pressure driven damage. This behavior results from the fact that, when $\Delta p_1$ increases with time, the mass flow on the damage-side of the damage front increases with time, and this increased flow drives the front forward much more quickly than when $\Delta p_1$ is constant, as in the pressure driven case.\par
\section{Surface flow rate for constant pressure injection}\label{FlowRates}
The fluid mass flow rate measured at the ground surface as a function of time is a commonly measured quantity in applications. We first derive an expression for the flow rate in the absence of damage. Afterward, we show how to obtain predicted flow rate for a spherical subsurface failure front geometry.\par
\subsection{Flow rate for the case of no damage}
Consider the case of fluid injection at constant pressure $p_H$ into a homogeneous medium of pressure $p_L < p_H$, and with no ensuing damage front. In this case, a pressure pulse spreads out radially from the injection point to approximately the radius $\sqrt{at}$ after a passage of time $t$. Using (\ref{SphericalGenSoln}), we approximate the pressure profile as
\begin{linenomath*} \begin{equation}
p(r) \approx p_H - \frac{\Delta p \sqrt{at}}{\sqrt{at}-r_0}\left(1 - \frac{r_0}{r}\right).
\end{equation} \end{linenomath*}
The pressure gradient near the injection point is thus
\begin{linenomath*} \begin{equation}
\frac{dp}{dr} \approx -\frac{\Delta p\sqrt{at}}{r_0(\sqrt{at}-r_0)}.
\end{equation} \end{linenomath*}
Neglecting gravitational effects and integrating the volumetric fluid flux over the surface of a sphere of fixed radius $r_0$ yields the flow rate 
\begin{linenomath*} \begin{equation} \label{NoDamage}
\mathcal{F}_0 \approx \frac{4\pi \rho k r_0 \Delta p \sqrt{at}}{\mu (\sqrt{at}-r_0)}.
\end{equation} \end{linenomath*}
Therefore, the flow rate is expected to approach the constant
\begin{linenomath*} \begin{equation} \label{NoDamageConstant}
\frac{4\pi\rho k r_0\Delta p}{\mu},
\end{equation} \end{linenomath*}
as $t \rightarrow \infty$. This formula also describes the flow rate for the case of ``full damage'', i.e., the situation that prevails after a damage front has progressed as far as possible and damage has ceased.
\subsection{Flow rate for the case of a spherical damage front}
We now consider the case of an over-pressure induced spherically shaped propagating damage front. Using equation (\ref{SphrPresLeft}) to calculate the volumetric flux at the injection well and integrating this flux over the surface of a sphere with radius $r_0$ yields
\begin{linenomath*} \begin{equation} \label{SphereFlowRate}
\mathcal{F}_{sph} = \frac{4\pi\rho k_1 \Delta p_1 r_0 R}{\mu(R-r_0)},
\end{equation} \end{linenomath*}
where $R$ is given by equation (\ref{SphrDamage}). This flow rate approaches (\ref{NoDamageConstant}) as $R \rightarrow \infty$, regar\begin{equation}
\frac{d}{dt}\sqrt{\kappa t} = \frac{1}{2}\sqrt{\frac{\kappa}{t}}
\end{equation}
where $\kappa$ is the thermal diffusivity, regardless of the particular form that $R$ takes. In the case of thermally induced damage, substituting (\ref{PressureDrop}) into (\ref{SphereFlowRate}) yields
\begin{linenomath*} \begin{equation} \label{SphereFlowRateTherm}
\mathcal{F}_{sph} = \frac{4\pi \rho k_1 \Delta p r_0 R}{\mu (R_{max}-r_0)},
\end{equation} \end{linenomath*}
where $R$ is now given by (\ref{ThermalSphrFront}). In this case we note that 
\begin{linenomath*} \begin{equation} \label{NoPorosDepend}
\frac{\partial \mathcal{F}_{sph}}{\partial \Delta \phi} \approx \frac{4\pi\rho k_1 \Delta p r_0}{\mu}\left( \frac{1}{R_{max}}\frac{\partial R}{\partial \Delta \phi} - \frac{R}{R_{max}^2}\frac{\partial R_{max}}{\partial \Delta \phi} \right).
\end{equation} \end{linenomath*}
However, if $R \gg r_0$ and $R_{max} \gg r_0$ then
\begin{linenomath*} \begin{equation}
\frac{1}{R_{max}}\frac{\partial R}{\partial \Delta \phi} \approx -\frac{R}{3 R_{max}\Delta\phi},
\end{equation} \end{linenomath*}
and
\begin{linenomath*} \begin{equation}
\frac{R}{R_{max}^2}\frac{\partial R_{max}}{\partial \Delta\phi} = -\frac{R}{3R_{max}\Delta\phi},
\end{equation} \end{linenomath*}
so that these terms in (\ref{NoPorosDepend}) exactly cancel one another. Hence, even though the position of the damage front depends on $\Delta \phi$, in the case of thermally driven damage the flow rate does not.
\section{Discussion}\label{Disc}
\subsection{The limit $\Delta \phi \rightarrow 0$}
Up until now we have assumed that, upon mechanical failure, the porosity increases. However, in some cases it is possible for the permeability to change by a large amount while the change in porosity is very small. It makes sense, then, to inquire into the possibility that the increase in porosity is zero or near zero; however, our formalism must be slightly altered in this case. For example, equation (\ref{ApproxLambda}) can be written
\begin{linenomath*} \begin{equation}
\Delta \phi \lambda^2 + \frac{2k_2 \Delta p_2}{\mu \sqrt{a_2}}\lambda - \frac{2k_1 \Delta p_1}{\mu}=0,
\end{equation} \end{linenomath*}
and in the limit $\Delta \phi \rightarrow 0$ the quadratic term vanishes. Hence, the correct formula in this case is not (\ref{ApproxLambda}) but
\begin{linenomath*} \begin{equation} \label{DeltaPorosZero}
\lambda \approx \frac{\sqrt{a_2}k_1\Delta p_1}{k_2 \Delta p_2}.
\end{equation} \end{linenomath*}
In the case of zero damage, i.e., when $p_D=p_H$, the above equation gives $\lambda = 0$ as expected.
\subsection{Variation in density due to movement of the front} \label{NeglectDens}
We have neglected variations in fluid density resulting from movement of the damage front, but we now show that these variations are negligible. For the same case as in section \ref{VertFront}, suppose that the damage front moves from position $x_1$ to $x_2$ over a small interval of time. Then the pressure at $x_1$ during this interval will have increased by amount
\begin{linenomath*} \begin{equation}
\Delta p \approx -\frac{\partial p_1}{\partial x}\bigg|_X \Delta X,
\end{equation} \end{linenomath*}
where $\Delta X \equiv x_2 - x_1$. This pressure increase, by (\ref{DensPres}), leads to an increase in density
\begin{linenomath*} \begin{equation}
\Delta \rho \approx -\frac{\partial p_1}{\partial x}\bigg|_X \Delta X \beta \rho_0.
\end{equation} \end{linenomath*}
Hence, the discrete form of condition (\ref{StefanCondHoriz}), when modified to include this density variation, is
\begin{linenomath*} \begin{equation} \label{StefanWithDens}
-\frac{k_1}{\mu} \frac{\partial p_1}{\partial x}\bigg|_{x=X} = -\frac{k_2}{\mu} \frac{\partial p_2}{\partial x}\bigg|_{x=X} + \Delta \phi \frac{\Delta X}{\Delta t}\left( 1 - \frac{\partial p_1}{\partial x}\bigg|_X\beta\Delta X \right).
\end{equation} \end{linenomath*}
In the limit as $\Delta t \rightarrow 0$, equation (\ref{StefanWithDens}) reduces to (\ref{StefanCondHoriz}).
\subsection{Field data comparison}
\begin{figure} \centering
\includegraphics[width=0.7\textwidth]{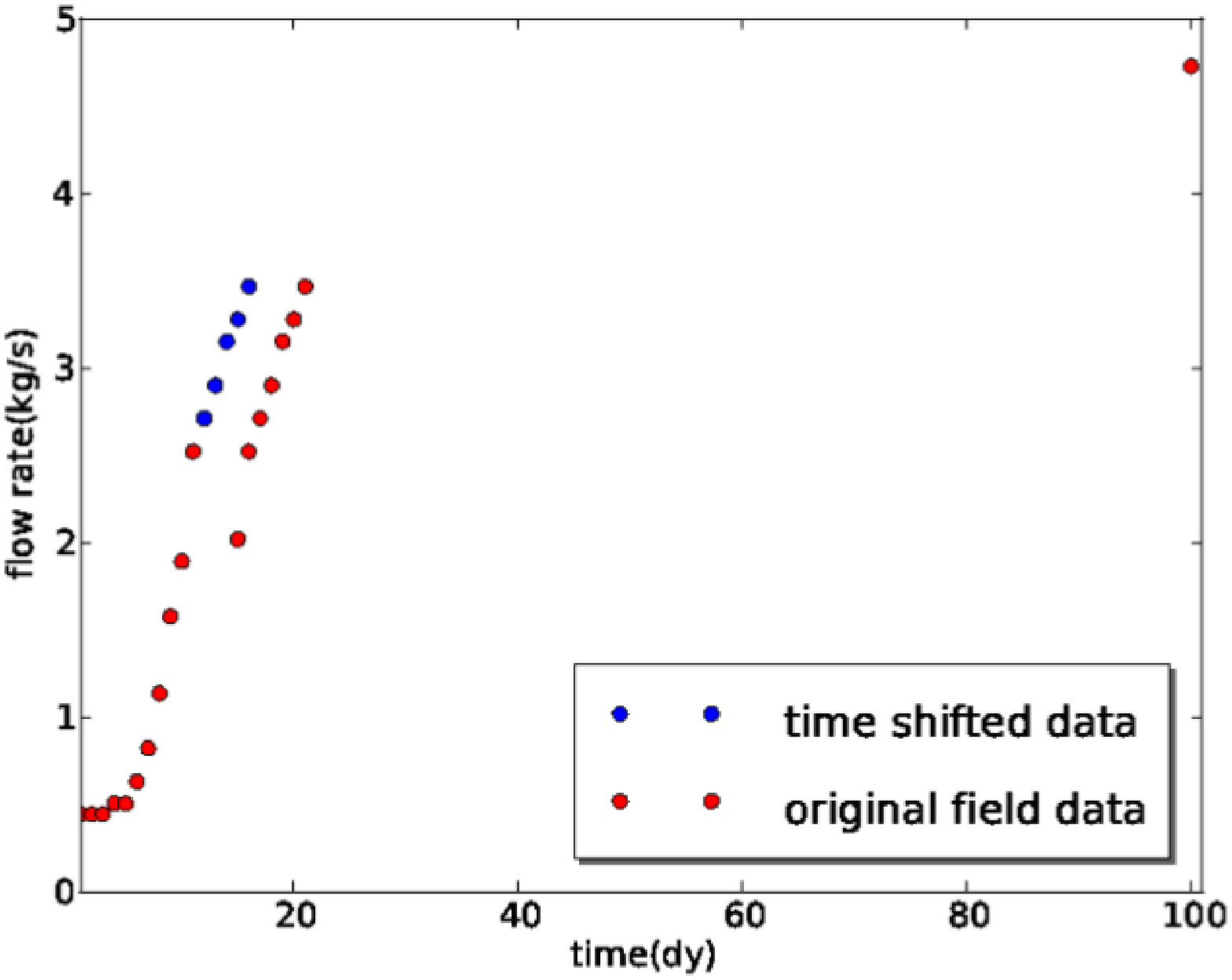}
\caption{Flow rates obtained by Chabora et al. \cite{Chabora} during constant pressure injection of fluid at $T\approx100^{\circ}$C into rock at $T \approx 190^{\circ}$C (red circles); data past eleven days shifted three days backward, and points overlapping with the earlier data removed (blue circles).} \label{ChaboraDataFig}
\end{figure}

\begin{figure} \centering
\includegraphics[width=0.7\textwidth]{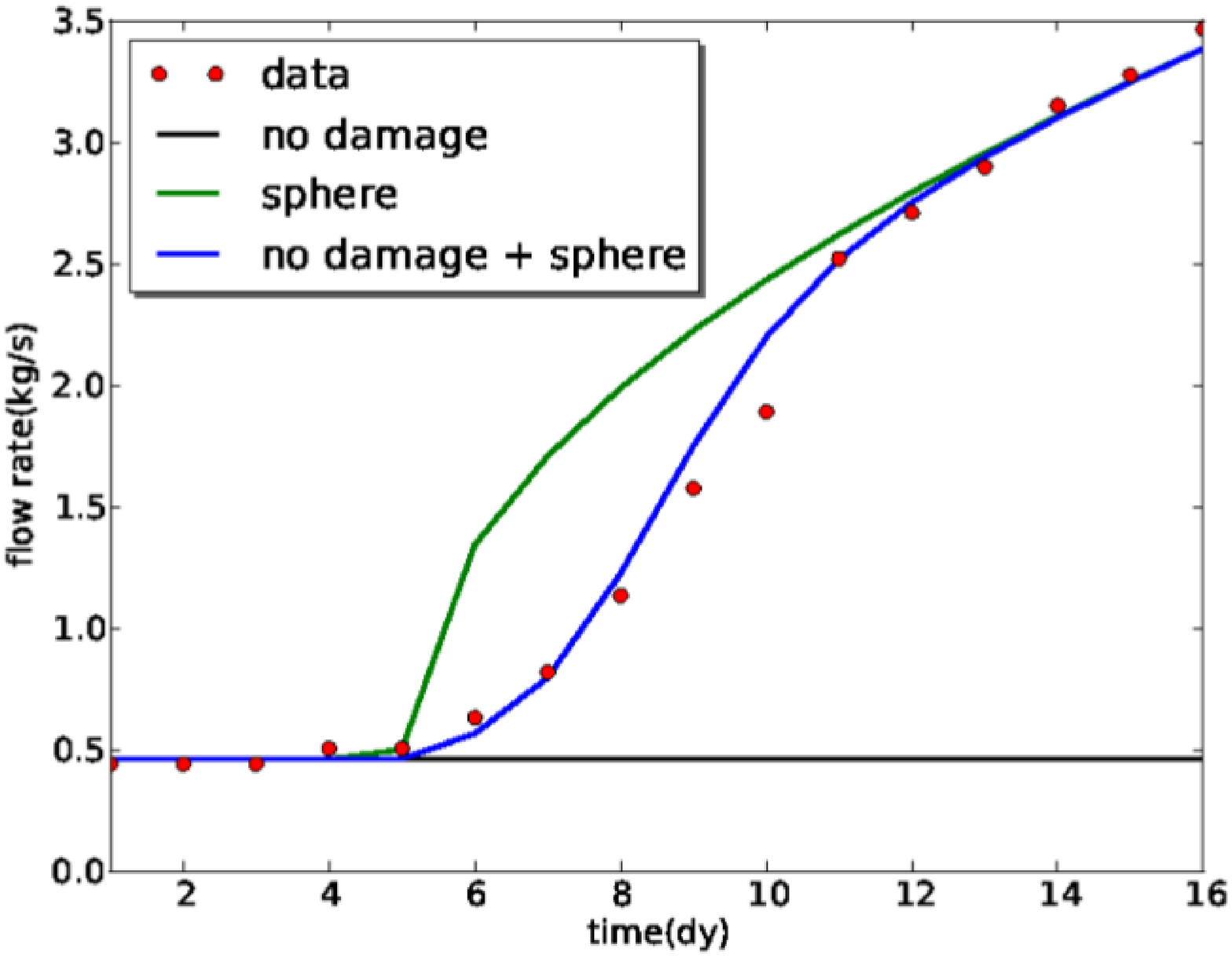}
\caption{Comparison between the model solution (\ref{ModelSoln}, blue), flow rates expected in the case of no damage (black), flow rates expected from an expanding spherical damage front starting at five days (green), and the Desert Peak field data (red dots).} \label{MatchAttemptFig}
\end{figure}

Chabora et al. \cite{Chabora} have reported surface flow rate data during constant pressure subsurface injection over the course of 100 days for a site near Desert Peak, NV (see Figure \ref{ChaboraDataFig}, red circles). Most of the data was gathered within the first 16 days, with one data point at 100 days. There is an interruption in data for about 3 days starting at day 12 due to pump failure. Because the resumed flow rate curve strongly resembles a continuation of the initial curve at a later time, and because the time during cessation of pumping is small relative to the thermal relaxation time of the reservoir, we align the resumed curve with the initial one as shown also in Figure \ref{ChaboraDataFig} (a ``continuous'' curve is depicted with red circles at earlier times and blue ones at later times). The resulting curve would likely have resulted if the pump had not failed. The temperature of the injected fluid was approximately 100$^\circ$C at the subsurface injection point, while the in situ temperature was approximately 190$^\circ$C. On the other hand, the injection pressure was 13.1 MPa, compared to an in situ pressure of about 9 MPa; therefore, rock failure was likely dominated by effects of thermal contraction. Consequently, we employ formula (\ref{SphereFlowRateTherm}) above to calculate flow rates associated with the rock failure. Because we are ignoring the effect of gravity, we attempt to match only the data from the first 16 days. The flow rate is approximately constant for the first several days, and according the formulae we have developed above, such a curve suggests that damage has no yet been initiated. We use a weight function to interpolate between the damage-free and spherical-damage-front solutions in the region where one regime begins to transition to the other. The weight function is taken as
\begin{linenomath*} \begin{equation} \label{WeightFunc}
w = 1 - \frac{1}{1+exp[{-b(t-t_{shf}-\Delta t)}]},
\end{equation} \end{linenomath*}
where $b$ is a constant with dimensions of inverse time, $t_{shf}$ is the time at which the failure geometry begins to transition from lack of damage to a spherical damage mode, and $\Delta t$ is the width of the region of overlap between these modes of failure. The total flow rate is thus
\begin{linenomath*} \begin{equation} \label{ModelSoln}
\mathcal{F} = w\mathcal{F}_0 + (1-w)\mathcal{F}_{sph}.
\end{equation} \end{linenomath*}
Table \ref{FieldCompare} gives the values of the parameters used to fit the field data. Figure \ref{MatchAttemptFig} shows the comparison between our model solution and the data. The blue curve is given by (\ref{ModelSoln}), the black curve is the predicted flow rate in the case of no damage, and the green curve shows the predicted flow rates for a spherical damage front only, all of these curves being calculated using the same model parameters. From equation (\ref{NoPorosDepend}), the choice of $\Delta \phi$ does not influence the flow rate. \par
When all observations are taken into account, the variables in our model are all constrained. The parameter $\Delta t$ is determined by noting that the observed flow rates transition from constant to non-constant flow rates over a span of a few days. The transition time, $t_{shf}$, occurs at about five days after injection begins. If $t_{shf}$ and $\Delta t$ are expressed in units of days, $b$ is then one inverse day, because it is a normalization factor converting time into the dimensionless time that is appropriate for the argument of an exponential function. The injection and in situ far-field pressures are constrained from observation and from the hydrostatic pressure profile, respectively. The fluid viscosity and density are determined from the average fluid pressure and temperature together with standard equations of state for pure water. The radius of the injection wellbore is known to be about $r_0 \approx 0.23$ m. For the pre-damage curve to match the flow rates at times less than five days, $k_2$ must be set equal to $10^{-14}$  m$^2$. For the flow rates corresponding to rock failure, $R_{max}$ is determined from equation (\ref{MaxDistance}) and noting that the observed flow rate curve appears to level off starting at about forty days, giving $t_{max} \approx$ 40 days (we are here assuming that the flattening of the curve starting at around this time corresponds to a transition to cessation of damage). We also note that, as $R_{max} \propto t_{max}^{1/3}$, the resulting flow rate is not very sensitive to the time chosen for $t_{max}$. The predicted curve corresponding to a spherical damage front (using (\ref{SphereFlowRateTherm})) can only match the observed flow rates past five days if $k_2$ is set to about $1.2 \times 10^{-13}$ m$^2$ - about twelve times the pre-damage permeability.\par
If we assume that the damage front coincides at all times with some isotherm - regardless of whether the temperature transitions smoothly from one side of the damage front to the other or whether, like the damage front, it has a sharp transition - then the velocity of the damage front can be used to estimate the relative strengths of diffusive versus total (advective plus diffusive) heat transfer. Assuming that the damage front is far from $r_0$ and differentiating equation (\ref{ThermalSphrFront}) with respect to time gives
\begin{linenomath*} \begin{equation} \label{VelocDamage}
\frac{dR}{dt} \approx \frac{1}{2}\sqrt{\frac{2k_1\Delta p r_0}{\mu \Delta \phi R_m t}}.
\end{equation} \end{linenomath*}
The velocity of a purely diffusive temperature front is roughly
\begin{linenomath*} \begin{equation} \label{DiffuseFront}
\frac{d}{dt}\sqrt{\kappa t} = \frac{1}{2}\sqrt{\frac{\kappa}{t}},
\end{equation} \end{linenomath*}
where $\kappa$ is the thermal diffusivity. Dividing \ref{VelocDamage} by \ref{DiffuseFront} yields the dimensionless number 
\begin{linenomath*} \begin{equation}
\chi \equiv \sqrt{\frac{2k_1\Delta p r_0}{\mu \Delta \phi R_m \kappa}}.
\end{equation} \end{linenomath*}
Assuming that $10^{-2} \le \Delta \phi \le 10^{-1}$ and using the values in Table \ref{FieldCompare}, we obtain the result that $16 \le \chi \le 36$. Hence, the model predicts that advection is very significant compared to diffusion for this system.\par
Finally we note that, although we have not included the effects of thermal expansion on the density in equation (\ref{SphericalMassConserve}), the error comitted is small as long as the fractional change in density is small. This is because the second term on the left side of the steady state mass balance equation
\begin{linenomath*} \begin{equation}
\nabla^2 p + \frac{\nabla \rho}{\rho}\cdot\nabla p = 0,
\end{equation} \end{linenomath*}
is small compared to the first if $\Delta \rho/\rho_0$ is small compared to unity. Even in the present case of thermally driven damage, the fractional change in density is only on the order of ten percent; therefore, the error committed in neglecting the density variation is acceptable for the purposes of this study.
\section{Conclusion}
Damage induced by fluid injection modifies subsurface permeabilities and porosities, causing both to be functions of pore pressure. Even though this dependence on pressure renders the mass balance equation nonlinear, we have been able to obtain a global analytic solution for the pore pressure in the case of a vertical propagating damage front front via an analogy with the classical Stefan problem, where a moving surface of discontinuous material properties splits the solution domain into two parts. A formula was derived stating that gravitational effects may be ignored for sufficiently small times. We have also derived approximate expressions for the position of the damage front as a function of time, which are valid for the cases of planar and spherical propagation front geometries. These expressions show that a thermally-induced damage front propagates much faster than one induced by over-pressure, for the same values assigned to the model parameters. Finally, using these expressions, we derived approximate formulae for the surface flow rates under constant pressure injection for the case of spherical damage front geometry. When compared to recorded flow rate data from a particular site near Desert Peak, NV, our model suggests one possible interpretation of the data is that subsurface failure began at about five days after commencement of fluid injection, transitioning from the pre-damage regime to one of spherical damage front geometry over the course of 16 days.

\section*{Acknowledgements}
This work was supported in part by the Department of Energy's Fossil Energy Program through the National Energy Technology Laboratory, and by the US DOE Office of Geothermal Technologies under Work Authorization No. GT-100036-12\_Revision 1, EERE agreement No. 25316. This support is greatly appreciated. Satish Karra thanks U.S. Department of Energy for the support through the geothermal project DE-EE0002766. The authors would also like to acknowledge insightful review and comments by David Dempsey that led to significant improvements.

\appendix
\renewcommand{\theequation}{A-\arabic{equation}}
% redefine the command that creates the equation no.
\setcounter{equation}{0}  % reset counter 
%\section*{APPENDIX}  % use *-form to suppress numbering
\section{Justification for neglecting $\nabla \rho$}
Substituting (\ref{VolFlux}), (\ref{PorosPres}), and (\ref{DensPres}) into (\ref{MassConserve1}) yields
\begin{linenomath*} \begin{equation} \label{FullMassBalance}
\widetilde{\beta}\frac{\partial p}{\partial t} - \frac{k}{\mu}\frac{\rho}{\rho_0}\nabla^2 p - \frac{2kg\rho\beta}{\mu}\nabla z \cdot \nabla p - \frac{\beta k}{\mu}(\nabla p)^2 = 0,
\end{equation} \end{linenomath*}
with
\begin{linenomath*} \begin{equation} \label{EffCompress}
\widetilde{\beta} \equiv \phi \beta + \frac{\rho}{\rho_0}\alpha. 
\end{equation} \end{linenomath*}
Dividing (\ref{FullMassBalance}) by the term proportional to $\nabla^2 p$ leads to the dimensionless equation
\begin{linenomath*} \begin{equation}
\frac{\rho_0}{\rho}\left( \frac{\widetilde{\beta} \mu}{k \nabla^2 p} \frac{\partial p}{\partial t} \right) - 1 - 2\beta\rho_0 g \frac{\nabla z \cdot \nabla p}{\nabla^2 p} - \frac{\rho_0}{\rho}\beta\frac{(\nabla p)^2}{\nabla^2 p} = 0.
\end{equation} \end{linenomath*}
Now consider a small vertical section of porous material of height $\Delta z$ over which the pressure varies by amount $\Delta p$, and suppose the time variation of $p$ over an interval of time $\Delta t$ is equal to $\xi \Delta p$ for some constant $\xi$. Then the first term on the left hand side is in order of magnitude
\begin{linenomath*} \begin{equation} \label{FirstTerm}
\frac{\xi\widetilde\beta\mu\Delta z^2}{k\Delta t},
\end{equation} \end{linenomath*}
where we have assumed that $\rho_0/\rho \approx 1$ and that the order of $\nabla^2 p$ is $\Delta p/\Delta z^2$. Term (\ref{FirstTerm}) is not in general small compared to unity. The third term has order of magnitude
\begin{linenomath*} \begin{equation}
2\beta\rho_0 g \Delta z,
\end{equation} \end{linenomath*}
and due to the smallness of $\beta$, only approaches unity for very large values of $\Delta z$. The fourth term varies as
\begin{linenomath*} \begin{equation}
\beta \Delta p,
\end{equation} \end{linenomath*}
and is small compared to unity except for very large values of $\Delta p$. Therefore, for the parameter regime of interest in this study, the dominant balance in equation (\ref{FullMassBalance}) is between the first and second terms on the left hand side.

\newpage
\begin{table}[ht]
\caption{Comparison of closed form and numerical values of $\lambda$} \label{LambdaCompare}
\centering
\begin{tabular}{c c c c c c c}
\hline\hline 
$k_1$(m$^2$) & $k_2$(m$^2$) & $\phi_1$ & $\phi_2$ & closed form $\lambda$ ($ms^{-1/2}$) & numerical $\lambda$ ($ms^{-1/2}$) & rel. error $\%$ \\ [0.5ex]
\hline \\ [-1ex]
10$^{-12}$ & 10$^{-14}$ & 0.101  & 0.1 & 2.116 & 2.109 & 0.3 \\ 
10$^{-12}$ & 10$^{-14}$ & 0.2    & 0.1 & 0.213 & 0.213 & 2$\times$10$^{-2}$ \\
10$^{-13}$ & 10$^{-14}$ & 0.1001 & 0.1 & 1.979 & 1.929 & 2.6 \\
10$^{-13}$ & 10$^{-14}$ & 0.3    & 0.1 & 4.759 $\times$10$^{-2}$ & 4.763 $\times$10$^{-2}$& 6.8$\times$10$^{-2}$ \\
10$^{-14}$ & 10$^{-16}$ & 0.15   & 0.1 & 3.012 $\times$10$^{-2}$ & 3.013$\times$10$^{-2}$ & 3.2$\times$10$^{-2}$ \\ 
10$^{-15}$ & 10$^{-16}$ & 0.152 & 0.15 & 4.671$\times$10$^{-2}$ & 4.691$\times$10$^{-2}$ & 0.4 \\ 
10$^{-12}$ & 10$^{-13}$ & 0.152 & 0.15 & 1.477 & 1.484 & 0.4 \\ [1ex]
\hline 
\end{tabular}
\label{table:lambdacompare}
\end{table}

\begin{table}[ht]
\caption{Parameters used to match the analytic solution to field data} \label{FieldCompare}
\centering 
\begin{tabular}{c c c c}
\hline\hline
Parameter & Value & Parameter & Value \\
\hline \\ [-1ex]
$k_1$ (m$^2$) & 1.2 $\times$ 10$^{-13}$ & $p_H$ (MPa) & 13.1 \\ 
$k_2$ (m$^2$) & 10$^{-14}$ & $p_L$ (MPa) & 9 \\
$\rho$ (kg/m$^3$) & 980 & $r_0$ (m) & 0.23 \\
$\mu$ (Pa$\cdot$s) & 0.25$\times$10$^{-3}$ & $b$ (days$^{-1}$) & 1 \\
$t_{shf}$ (days) & 5 & $\Delta t$ (days) & 3 \\
$t_{max}$ (days) & 40  &  &    \\ [1ex]
\hline 
\end{tabular}
\label{table:fieldcompare}
\end{table}

\newpage

\end{document}